\documentclass[aps,pre,twocolumn]{revtex4}
\usepackage{amsmath}
\usepackage{amssymb}
\usepackage{natbib}
\usepackage{color}
\usepackage{graphicx}
\usepackage{subfigure}

\begin{document}

\bibliographystyle{apsrev}

\title{Extreme bendability of DNA double helix due to bending asymmetry}

\author{H. Salari}
\address{Sharif University of
Technology, Department of Physics, P.O. Box 11365-8639, Tehran,
Iran.}
\author{B. Eslami-Mossallam}
\address{Sharif University of
Technology, Department of Physics, P.O. Box 11365-8639, Tehran,
Iran.}

\author{M.S. Naderi}
\address{Sharif University of
Technology, Department of Physics, P.O. Box 11365-8639, Tehran,
Iran.}

\author{M.R. Ejtehadi}
\email{ejtehadi@sharif.edu}
\address{Sharif University of
Technology, Department of Physics, P.O. Box 11365-8639, Tehran,
Iran.}

\date{\today}

\begin{abstract}
 Experimental data of the DNA cyclization (J-factor) at short length scales, as a way to study the elastic behavior of tightly bent DNA, exceed the theoretical expectation based on the wormlike chain (WLC) model by several orders of magnitude. Here, we propose that asymmetric bending rigidity of the double helix in the groove direction can be responsible for extreme bendability of DNA at short length scales and it also facilitates DNA loop formation at these lengths. To account for the bending asymmetry, we consider the asymmetric elastic rod (AER) model which has been introduced and parametrized in an earlier study [Eslami-Mossallam, B.; Ejtehadi, M. R. \textit{Phys. Rev. E} \textbf{2009}, \textit{80}, 011919]. Exploiting a coarse grained representation of DNA molecule at base pair (bp) level, and using the Monte Carlo simulation method in combination with the umbrella sampling technique, we calculate the loop formation probability of DNA in the AER model. We show that, for DNA molecule has a larger J-factor compared to the WLC model which is in excellent agreement with recent experimental data.
\end{abstract}

\maketitle
\section{Introduction}
Studying the elastic behavior of DNA molecules is important for understanding its biological functions. One of the most popular theoretical models to explain the elastic behavior of DNA is the harmonic elastic rod model, also called the wormlike chain (WLC) model \cite{Marko03, Towles}. In this model it is assumed that the elastic energy is a harmonic function of local deformations. The
WLC model can predict very accurately the elastic
properties of long DNA molecules and yielding a persistence length of about $50\,\mathrm{nm}$ for DNA \cite{Marko01, Towles}. However, recent experimental data suggest that, short DNA molecules are much more flexible than what is predicted by the WLC model \cite{Widom01, Widom02, Wiggins01, Yuan, Vafabakhsh}. For example, loop formation probability, i.e. the J-factor~\cite{Shimada}, for DNA molecules shorter than $100\,\mathrm{bp}$ ($\sim\,34\,\mathrm{nm}$) is several orders of magnitude higher than the prediction of the WLC model \cite{Widom01, Widom02, Vafabakhsh, Vologodskii01}. 

Different experimental procedures have been used to measure the cyclization probability for short DNA molecules. For example, in Cloutier and Widom's work~\cite{Widom02} the DNA molecules have short sticky ends. Therefore, when the two DNA ends are close to each other torsional and axial alignment are required to form a DNA loop, which is then stabilized by the ligase. Thus the J-factor depends on the concentration of the ligase in the experiment \cite{vologod2013}. On the other hand, Vafabakhsh and Ha have used DNA molecules with long sticky ends \cite{Vafabakhsh}. In this case it is expected that the rate of loop formation depends only on the probability that the two DNA ends reach together, and thus is directly related to DNA elasticity. In the both experiments the persistence length of short DNA molecules appear to be much shorter then $50\,\mathrm{nm}$.

It has been suggested that the anomalous elastic behavior of short DNA molecules is a consequence of formation low-energy kinks in highly bent DNA molecules~\cite{vologod2013, Nelson01, Wiggins02}. Also it has shown that local DNA melting of the cyclized DNA increases the J-factor at short length scales~\cite{Marko02,Menon2005}. Also, there has been efforts to explain the effect by introducing more structural details to elastic model (e.g., considering cooperativty~\cite{Menon2013,Xu2014} or anisotropy~\cite{Behrouz,Farshid2008} in bending rigidity of DNA). But these efforts were not successful, as it has been shown that anisotropy has no significant effect in these dimensions~\cite{Nelson02} and even leads to the stiffening of DNA if the molecule is confined in a two dimensional surface~\cite{Golnoosh}. The effect of electrostatic interaction has been studied which can increase the j-factor~\cite{Cherstvy2011}, but in this study we assume the DNA molecule is neutral and the electrostatic interaction would be ignored.

The DNA molecule in its B form suggests that DNA has an asymmetric structure, in the sense that the opposite grooves of the DNA are not equal in size and the structure
\cite{Calladine01}. Thus, one expects that the energy required to
bend the DNA is not only anisotropic, but also asymmetric, i.e. the energy required to bend the DNA over its major groove is not equal to the energy needed to bend over its minor groove to the same angle. There are
theoretical analysis \cite{Crick}, experimental evidences
\cite{Richmond}, and simulation studies \cite{Lankas01} which
show that the bending asymmetry may affect the elastic behaviour of DNA. In previous work we have introduced the
asymmetric elastic rod (AER) model to account for
the asymmetric bending of DNA~\cite{Behrouz2}.
In this paper, we evaluate the elastic properties of the AER model, namely the DNA looping probability, to reveal the relevance of the asymmetric bending for short extremely bent DNA molecules. To this end we exploit the Monte Carlo (MC) simulation in combination with the umbrella sampling (US) technique, which enables us to efficiently sample the rare cyclization events. We show that the AER model is in an excellent agreement with the experimental data of the J-factor at short length scales.

\section{Model and Method}
\subsection{Asymmetric elastic rod model} 
\begin{figure}[thb]
\begin{center}
\includegraphics[scale=0.25]{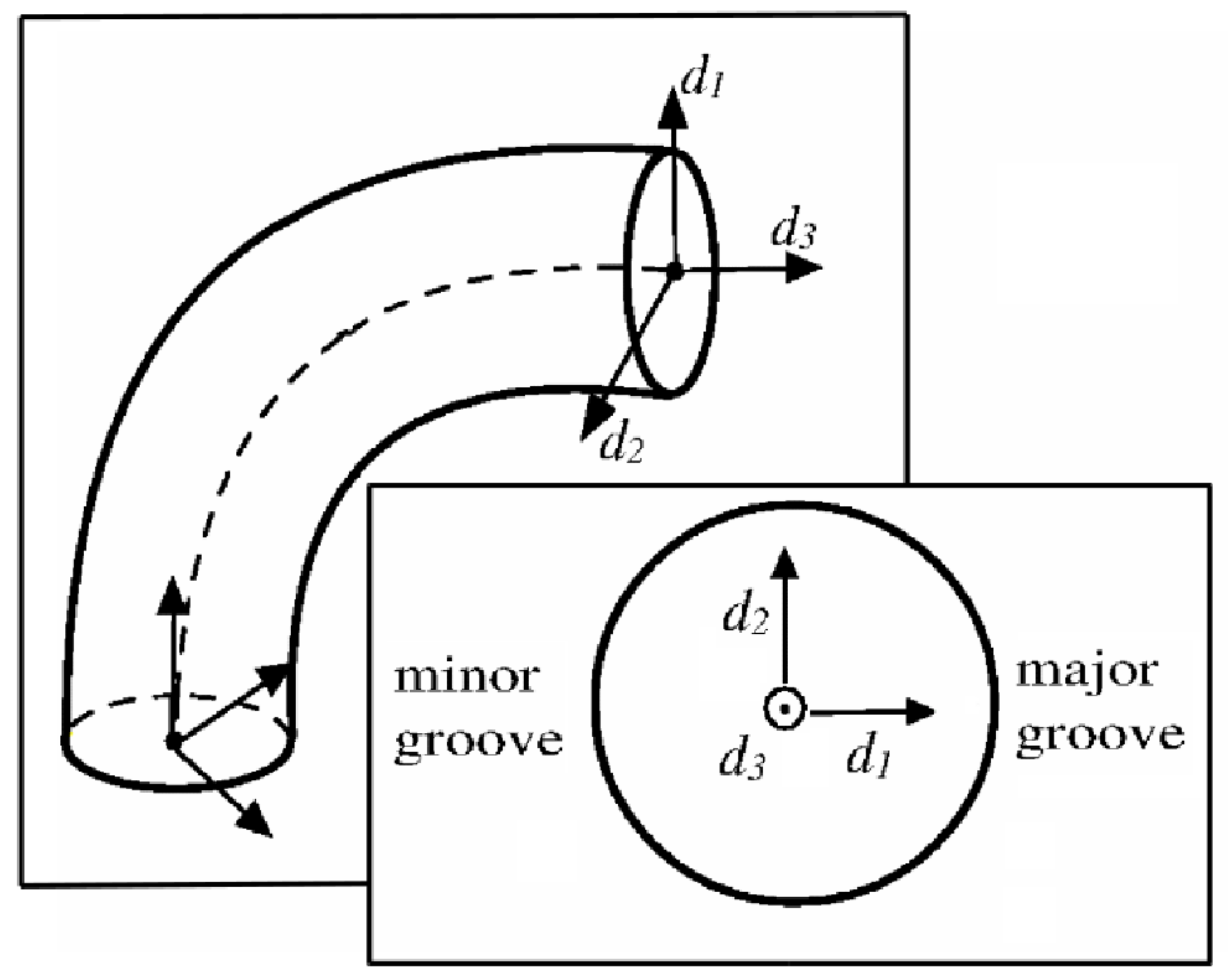}
\end{center}
\caption{Parametrization of the elastic rod. The local frame
$\{\hat{d_{1}}, \hat{d_{2}}, \hat{d_{3}}\}$ is attached to the
rod. \label{fig:1}}
\end{figure}

In the elastic rod model, a DNA is represented by a flexible
inextensible rod \cite{Marko03, Towles}, which can be deformed in
response of the external forces or torques. Here we use the
discrete elastic rod model \cite{Towles,Mergell}, where the rod is
discretized into segments each representing a DNA base pair. In
this model, the internal degrees of freedom of the base pairs are
neglected, and each base pair is considered as a rigid body. A
local coordinate  system (material frame) with an orthonormal basis $\{\hat{d_{1}},
\hat{d_{2}}, \hat{d_{3}}\}$ is attached to each base pair. As
depicted in Figure \ref{fig:1}, $\hat{d_{3}}$ is perpendicular to
the base pair surface, $\hat{d_{1}}$ lies in the base pair plain
and points toward the major groove, and $\hat{d_{2}}$ is defined
as $\hat{d_{2}}=\hat{d_{3}}\times \hat{d_{1}}$. Since it is
assumed that the DNA is inextensible, each base pair only has
three rotational degrees of freedom, and the position of the $(k+1)$th base pair with respect to the
$k$th base pair is denoted by the vector $\vec{r}(k)$ which is
given by \cite{Towles}
\begin{equation}
\label{position}\vec{r}(k)=l_0\,\hat{d_{3}}(k)\,,
\end{equation}
where $l_0=0.34\,\mathrm{nm}$ is the base pair separation.
The orientation of the
$(k+1)$th base pair with respect to the $k$th base pair is
determined by a rotation transformation $\mathbf{R}(k)$,
which can be parametrized by a vector
$\vec{\Theta}(k)$. The direction of $\vec{\Theta}(k)$ is normal to the plain of rotation of the $k$th base pair,
and its magnitude determines the rotation angle. The components of
$\vec{\Theta}(k)$ in the local coordinate system attached to the
$k$th base pair are denoted by $\Theta_1(k)$, $\Theta_2(k)$, and
$\Theta_3(k)$. These components can be regarded as three
rotational degrees of freedom of the base pairs around $\hat{d_{1}},
\hat{d_{2}}$ and $\hat{d_{3}}$, and are called
tilt, roll, and twist respectively \cite{Mergell}. If the values of these three
angles are known for all base pairs, the conformation of the DNA
can be uniquely determined.

For an inextensible DNA with $N$ base pair steps, the elastic
energy depends on the spatial
angular velocity $\vec{\Omega}(k)=\frac{\vec{\Theta}(k)}{l_0}$, then the elastic energy of the AER model \cite{Behrouz2} can be written as 
\begin{equation}
\label{LocalE} E=\sum_{k=1}^N
l_0\,\mathcal{E}^{\mathrm{\,AS}}_k[\vec{\Omega}(k)]\,,
\end{equation}
with
\begin{eqnarray}
\label{LocalAsymmetricE}\mathcal{E}^{\mathrm{\,AS}}_k\,[\vec{\Omega}(k)]=
\qquad\qquad\qquad\qquad\qquad\qquad\qquad\qquad\nonumber\\
k_\mathrm{B}T\Bigg[\frac{1}{2}A_1\,\Omega_1(k)^2+\frac{1}{2}A_2\,\Omega_2(k)^2+
\qquad \qquad \nonumber\\
\frac{1}{2}C\,(\Omega_3(k)-\omega_0)^2
+\,\frac{1}{3\,!}\,F^{2}\,\Omega_2(k)^3+
\ \nonumber\\
\frac{1}{4\,!}\,G^{3}\left(\Omega_1(k)^4+\Omega_2(k)^4\right)
\Bigg],
\qquad\qquad\qquad
\end{eqnarray}
where $k_\mathrm{B}$ is the Boltzmann constant and $T$
is the absolute temperature. The first three terms in equation
(\ref{LocalAsymmetricE}) correspond to the harmonic part of the elastic energy, which also appear in the WLC model. $A_1$, $A_2$ and
$C$ are the harmonic elastic constants of
DNA, and $\omega_0$ is its intrinsic twist.
\begin{figure}[thb]
\begin{center}
\includegraphics[scale=0.5]{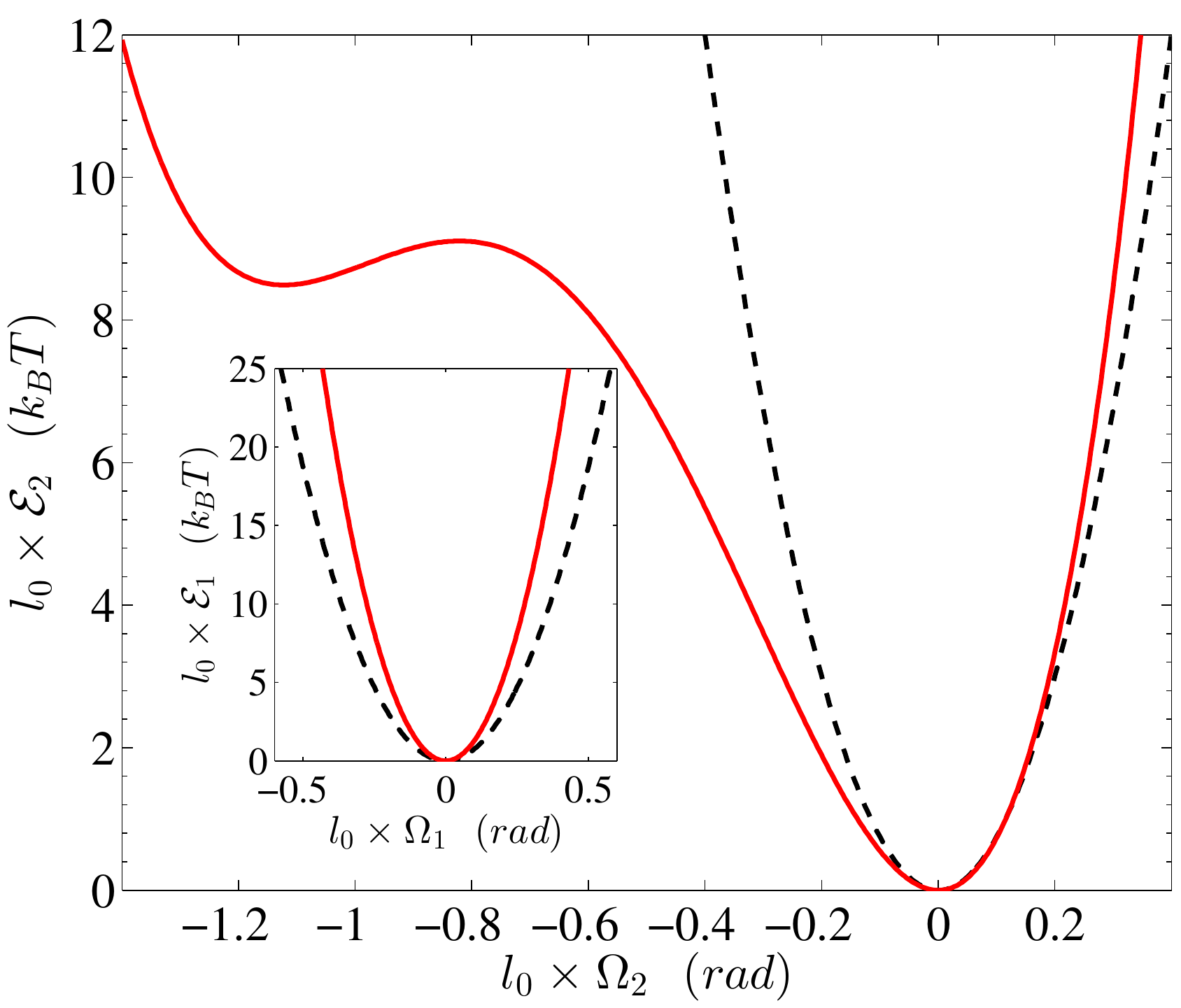}
\end{center}
\caption{(Color online) $\mathcal{E}_2$ as a function of
$\Omega_2$ for two different models, ``W" and ``A" (Table \ref{table}). The symmetric model ``W", dashed (black) curve, has one minimum and its curvature is positive everywhere.  The asymmetric model ``A", solid  (red) curve, has two minima.  $\mathcal{E}_1$ as a function of
$\Omega_1$ for both the models remain always convex (inset). \label{fig:2}}
\end{figure}
The remaining terms in equation (\ref{LocalAsymmetricE}) constitute the anharmonic
parts of the elastic energy. The term
$+1/3!\,F^2\,\Omega_2^3$ accounts for the asymmetric structure of DNA in the bending
energy, while the term $\frac{1}{4\,!}\,G^{3}\left(\Omega_1(k)^4+\Omega_2(k)^4\right)$ preserves the
stability and the consistency of the model.

Since there is no coupling term in the model, roll, tilt, and
twist can be regarded as independent deformations, and the energy
density can be decomposed into three separate terms
\begin{equation}
\label{Energy-decomposed}
\mathcal{E}^{\mathrm{AS}}[\Omega_1,\Omega_2,\Omega_3]=\mathcal{E}^{\mathrm{AS}}_1[\Omega_1]+\mathcal{E}^{\mathrm{AS}}_2[\Omega_2]+\mathcal{E}^{\mathrm{AS}}_3[\Omega_3]\,,
\end{equation}
where
\begin{equation}
\label{E1} \mathcal{E}^{\mathrm{AS}}_{1}[\Omega_1]=
k_\mathrm{B}T\Big[\frac{1}{2}A_1\,\Omega_1^2+
\frac{1}{4\,!}\,G^{\,3}\,\Omega_1^4\Big],\qquad\qquad\qquad
\end{equation}
\begin{equation}
\label{E2} \mathcal{E}^{\mathrm{AS}}_{2}[\Omega_2]=
k_\mathrm{B}T\Big[\frac{1}{2}A_2\,\Omega_2^2+\,\frac{1}{3\,!}\,F^{\,2}\,\Omega_2^3+
\frac{1}{4\,!}\,G^{\,3}\,\Omega_2^4\Big],\,\,
\end{equation}
and
\begin{equation}
\label{E3} \mathcal{E}^{\mathrm{AS}}_{3}[\Omega_3]=
\frac{1}{2}k_\mathrm{B}T\,
C(\Omega_3-\omega_0)^2\,.\qquad\qquad\qquad\qquad\quad
\end{equation}
Here, we use  parameters of the model~\cite{Behrouz2}, which is obtained by fitting the AER model to the experimental data of Wiggins \emph{et al.}~\cite{Wiggins01} (see the first row of Table \ref{table} as model ``A"). In this parametrization, the bending anisotropy also is considered, where $A_1 \neq A_2$. With this parametrization the roll energy, $\mathcal{E}_{2}$, has two minima (solid, red curve in Figure \ref{fig:2}); one at $\Omega_2=0$ and the second one  $\Omega_2=-3.3 \,\mathrm{nm}^{-1}$ corresponds to a negative roll of about $60^{\circ}$, with a roll energy about $8\,k_\mathrm{B}T$. The energy barrier between two minima is about $9\,k_{\rm B}T$. The existence of a second minimum in
$\mathcal{E}_2$ can lead to the formation of kinks in the minor
groove direction. With a large energy barrier between the two minima, one expects that the kinks rarely form in a free DNA at room temperature. However, if the DNA is forced to adopt a tightly bent conformation the probability of kink formation increases significantly.

The possibility of kink formation in the DNA structure has been considered previously by other authors. An atomistic structure for a
kinked DNA has been proposed by Crick and Klug~\cite{Crick}, who suggest that DNA can form a kink in the minor groove direction. Also, molecular dynamics simulations on a $94\,\mathrm{bp}$
minicircle \cite{Lankas01} show that kinks are formed, with
the same structure predicted by Crick and Klug. A simple model has been presented by Nelson, Wiggins, and Phillips
to describe the elasticity of kinkable elastic rods \cite{Nelson01}. This
model is mathematically equivalent to the models of local DNA melting \cite{Marko02, Menon2005, Palmeri}. Recently, Vologodskii and Frank-Kamenetskii have proposed another model for the kink formation in DNA~\cite{vologod2013}. In all of these models, the kinks are isotropic, i.e. they can be formed in any direction with equal probability.
On the contrary, in the AER model there is a privileged direction for the kink formation, i.e. the groove direction.

In order to compare the AER model with the WLC model, we also use another set of parameters here, which
are given in second row of Table \ref{table} as model ``W". As we will show in result section, at long length scales, these two models are equivalent and they yield the same persistence length, $l_p = 51 nm$. The tilt and roll energies, $\mathcal{E}_1$ and $\mathcal{E}_2$, of these two models are compared with each other in Figure \ref{fig:2}.
\begin{table}[hbt]
\begin{center}
\begin{tabular}[c]{|c|c|c|c|c|c|c|c|}
\hline
model & $A_1\,(\mathrm{nm})$ & $A_2\,(\mathrm{nm})$ & $F\,(\mathrm{nm})$ & $G\,(\mathrm{nm})$ & $C\,(\mathrm{nm})$ & $\omega_0\,(1/{\mathrm{nm}})$ \\
\hline {A} & {$87.00$} & {$43.50$} & {$7.90$} & {$3.20$} & {$100$} & {1.8} \\
\hline {W} & {$51.00$} & {$51.00$} & {$0.0$} & {$0.0$} & {$100$} & {1.8} \\
\hline
\end{tabular}
\end{center}
\caption{ Two different parameter sets for the AER and WLC models, indicated as models ``A" and ``W" respectively in this study. The parameter set ``A" is obtained from the fitting of
the model to the experimental data of Wiggins \emph{et al.}
\cite{Behrouz2} and its effective persitence length is about $51\,\mathrm{nm}$, same as molde ``W". \label{table}}
\end{table}
\subsection{{Calculation of J-factor} \label{US}}
\begin{figure}[thb]
\begin{center}
\includegraphics[scale=0.3]{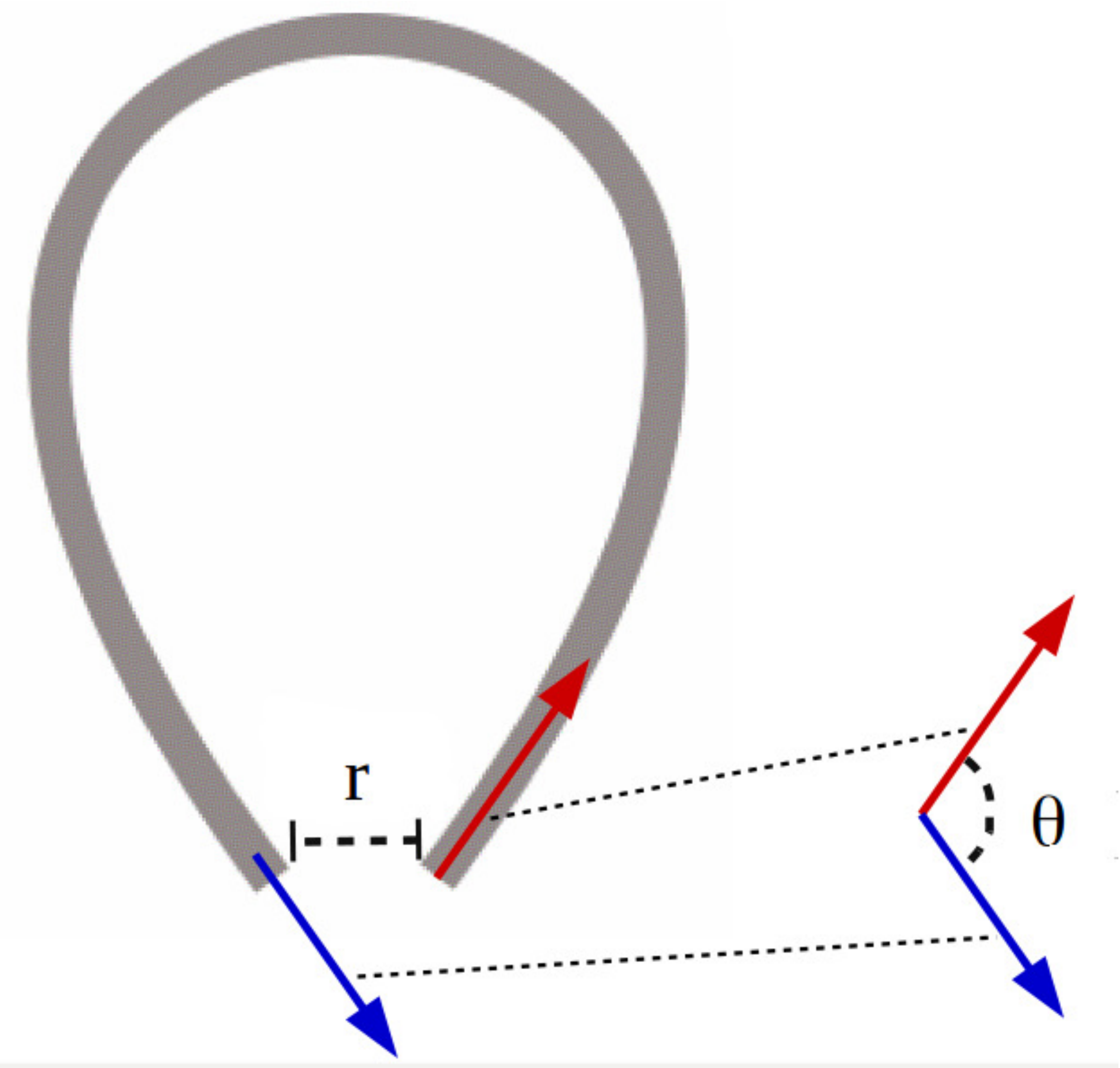}
\end{center}
\caption{(Color online) Schematic figure of a chain with terminal tangent vectors, $\hat{d}_3(0)$ and $\hat{d}_3(L)$, which are indicated by red and blue arrows, respectively. The end-to-end distance is shown by $r$, and $\theta$ is the angle between the tangent vectors.  \label{loopsch}}
\end{figure}

The loop formation probability of a polymer, which is known as J-factor \cite{Flory1}, is defined as the probability
that the two ends of the polymer meet each other with axial and torsional alignment. For simplicity we neglect the torsional alignment and only require that the two ends are close to each other while the two terminal tangent vectors are parallel. Denoting the separation between the two ends by $r$, and the angle between the two terminal tangent vectors by $\theta=\cos^{-1}(\hat{d}_3(0)\cdot\hat{d}_3(L))$ (see figure \ref{loopsch}), the J-factor for a DNA of length $L$ in molar unit is given by \cite{Flory1,Vologodskii05}
\begin{eqnarray}
\label{Jfactor}
J(L)=\lim_{\substack{r_0 \rightarrow 0 \\ \gamma_0\rightarrow 1}} \frac{1}{N_a}\frac{3}{4\pi r_0^3}\int_0^{r_0} K(r,L)\,\mathrm{d}r \
\qquad\qquad\nonumber\\
\times\, \frac{2}{1-\gamma_0}\int_{\gamma_0}^1 P_{\gamma}(\gamma ,L)\,\mathrm{d}\gamma ,
\end{eqnarray}
where $\gamma=\cos\theta$, $N_a$ is Avogadro's Number, $K(r,L)$ is the normalized distribution function of $r$, and $P_{\gamma}(\gamma,L)$ is the normalized distribution function of $\gamma $ under condition of $r=0$. 

The bending energy of DNA depends on the bending direction. Thus there is a implicit bend-twist coupling in this model. However this coupling can barely affect the J-factor if the DNA length is much larger than the helical pitch. Therefore we expect that for both models, the average trend of the J-factor is given by the equation~(\ref{Jfactor}). For end distances near zero ($r/L \approx 0$), the radial distribution function, $K(r,L)$, is proportional to $r^2$
\begin{equation}
\label{greenfunc}
K(r,L) \longrightarrow  4\pi r^2K_0(L) \qquad\qquad \mathrm{for}\,\,\, r/L\longrightarrow0,
\end{equation}
where $K_0(L)$ is a $r$-independent function~\cite{Mehraeen2008a}. In addition we have
\begin{equation}
\lim_{\gamma_0\rightarrow 1}\frac{1}{1-\gamma_0}\int_{\gamma_0}^1 P_{\gamma}(\gamma ,L)\,\mathrm{d}\gamma\, = \, P_{\gamma}(\gamma=1,L),
\label{k0}
\end{equation}
then the equation (\ref{Jfactor}) can be written as
\begin{eqnarray}
\label{JfactorG}
J(L) = J_0(L)\times2P_{\gamma}(1,L)
\end{eqnarray}
where
\begin{eqnarray}
\label{J0factorG}
J_0(L) = \frac{1}{N_a}K_0(L),
\end{eqnarray}
is the unconstrained J-factor which dose not involve axial and torsional alignment.
\subsection{Simulations \label{sim}}
We exploited a Metropolis Monte Carlo (MC) simulation to evaluate the statistical properties of DNA. We do not include the self avoiding in the simulations, since the probability of self crossing is small for the short simulated DNA molecules.
For short DNA molecules the
loop formation probability is very low, and thus the DNA cyclization events are too rare to be observed in the simulations. To overcome this problem,
we used the method of Umbrella sampling (US)~\cite{US} to
evaluate the distribution functions $K(r,L)$ and $P_{\gamma}(\gamma , L)$.
To calculate $K(r,L)$ the reaction coordinate is the end-to-end distance, $r$, we divided $r$ into 100
successive windows, and for each window performed a separate MC
simulation in which a harmonic bias potential is applied to
the end-to-end distance of the DNA. All the harmonic potential have a common spring constant $\mathcal{A}^b_r=0.3\,k_\mathrm{B}T/\mathrm{nm}^2$, and the minimum of each potential lies at the center of the corresponding window. We then found the biased distribution for each individual simulation and used the Weighted
Histogram Analysis Method (WHAM) \cite{WHAM} to reconstruct the
unbiased distribution function. 
To calculate $P_{\gamma}(\gamma,L)$, we set the end-to-end distance to zero, then perform another US by dividing the range of variation of $\gamma $ into 100 windows and applying a harmonic bias potential with spring constant of $\mathcal{A}^b_{\gamma}=40\,k_\mathrm{B}T$, in each window. The unbiased distribution
function $P_{\gamma}(\gamma ,L)$ is then found by WHAM.

In each individual simulation during the umbrella sampling procedure, the first $10^5$ MC steps were disregarded to ensure the equilibration of the system and the next $2\times10^6$ MC steps were considered for sampling. We perform 5 independent simulations in each window to estimate errorbars. 
\begin{figure}[thb]
\begin{center}
\includegraphics[scale=0.45]{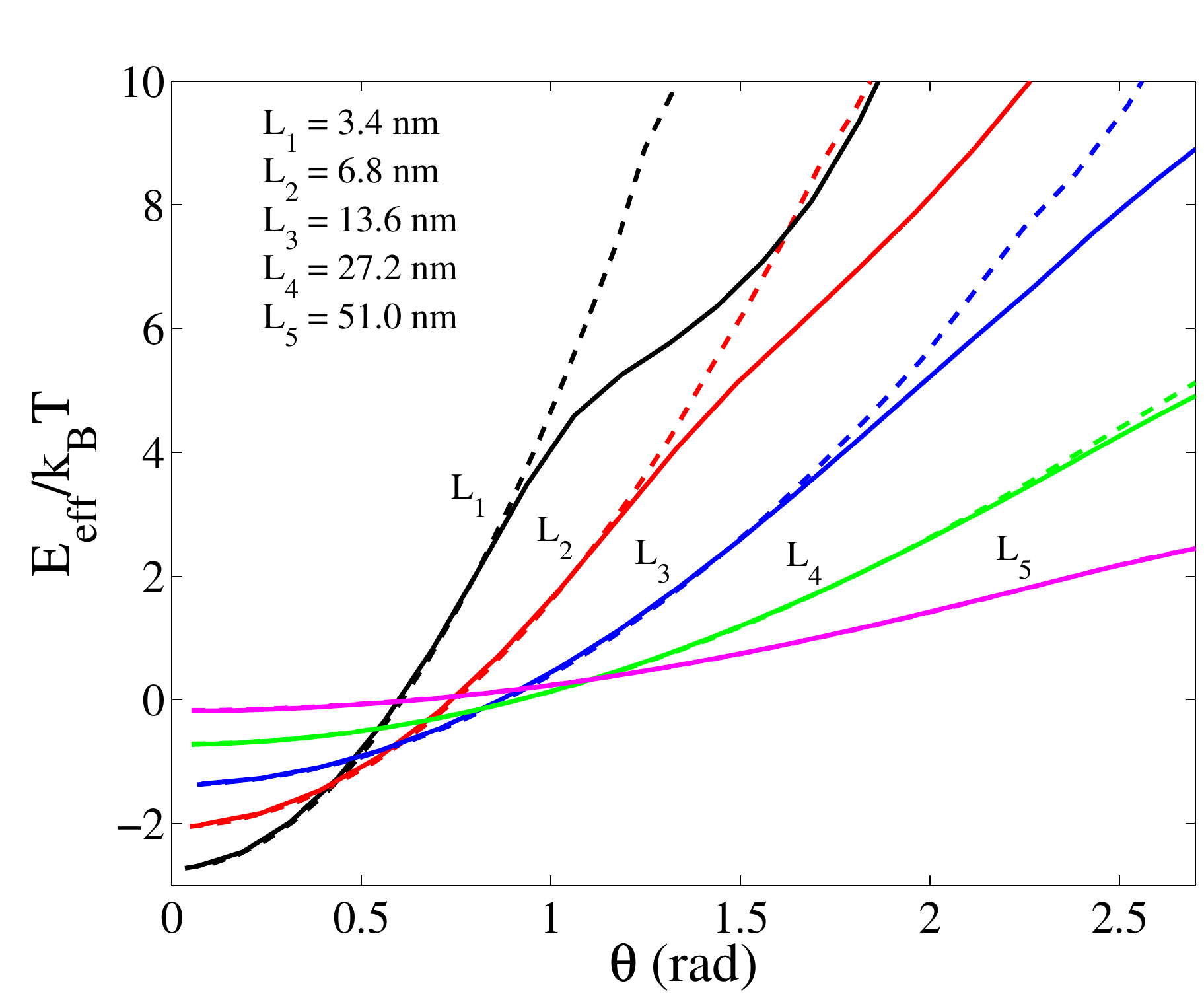}
\end{center}
\caption{(Color online) Monte Carlo simulation results of the effective bending energy, $E_{\mathrm{eff}}$, as a function of
the bending angle, $\theta$, for various chain lengths $L=3.4\,\mathrm{nm}$ (black), $L=6.8\,\mathrm{nm}$ (red), $L=13.6\,\mathrm{nm}$ (blue), $L=27.2\,\mathrm{nm}$ (green) and $L=51.0\,\mathrm{nm}$ (magenta). Solid and dashed curves correspond to ``A" and ``W" models, respectively (see Table \ref{table}).\label{fig:Eeff1}}
\end{figure}
\section{Results and Discussion}
\subsection{Persistence Length and Effective Bending Energy}
For a long DNA of length $L$, the persistence length, $l_p$, is defined as~\cite{Wiggins02}.
\begin{equation}
\label{correlation}
\langle \cos (\theta)\rangle = \exp(-L/l_p)\,.
\end{equation}
With the parameters given in Table \ref{table}, the asymmetric model (model ``A") and the wormlike chain model (model ``W") have a common persistence length of about $51\,\mathrm{nm}$. This means that for long DNA molecules with small deformations the two model are equivalent, and thus the asymmetric model is effectively reduced to an isotropic wormlike chain model. On the other hand, at large bending angles, one expects that the asymmetric structure of DNA affects its elastic properties. To show this, we evaluated the effective bending energy as a function of the bending angle $\theta$, which is defined as~\cite{Curuksu}
\begin{equation}
\label{Eeff}E_{\mathrm{eff}}(\theta ,L)=-k_\mathrm{B}T\,\ln\frac{P(\theta
,L) }{\sin \theta},
\end{equation}
where $P(\theta ,L)$ is the distribution function of the bending angle $\theta$ of a DNA of length $L$. Figure \ref{fig:Eeff1} compares the effective energies of the AER model (solid curves) with the WLC model (dashed curves) for different DNA lengths. One can see that, at small bending angles, both models follow a common parabola. However, at large bending angles, the effective bending energy of the asymmetric model falls beneath the parabola, which leads to extreme bendability of DNA or formation of kinks~\cite{Behrouz2}. The effect is suppressed as the DNA length increases. It is well expected  that, for long enough DNA, the effective energy  is independent on the structural details and it converges to a
parabola \cite{Wiggins01}.

As Figure \ref{fig:Eeff1} shows, the transition between the harmonic and non-harmonic region is smooth. This is because in the AER model DNA preserves its resistance against bending even in kink conformation. In simpler versions of kinkable elastic rod models~\cite{Nelson01,Wiggins02}, where the kinks are assumed to be completely flexible, there is a sharp transition in the curve of the effective bending energy between a parabola and a straight line with zero slope \cite{Nelson01}.
\begin{figure}[thb]
\begin{center}
\includegraphics[scale=0.8]{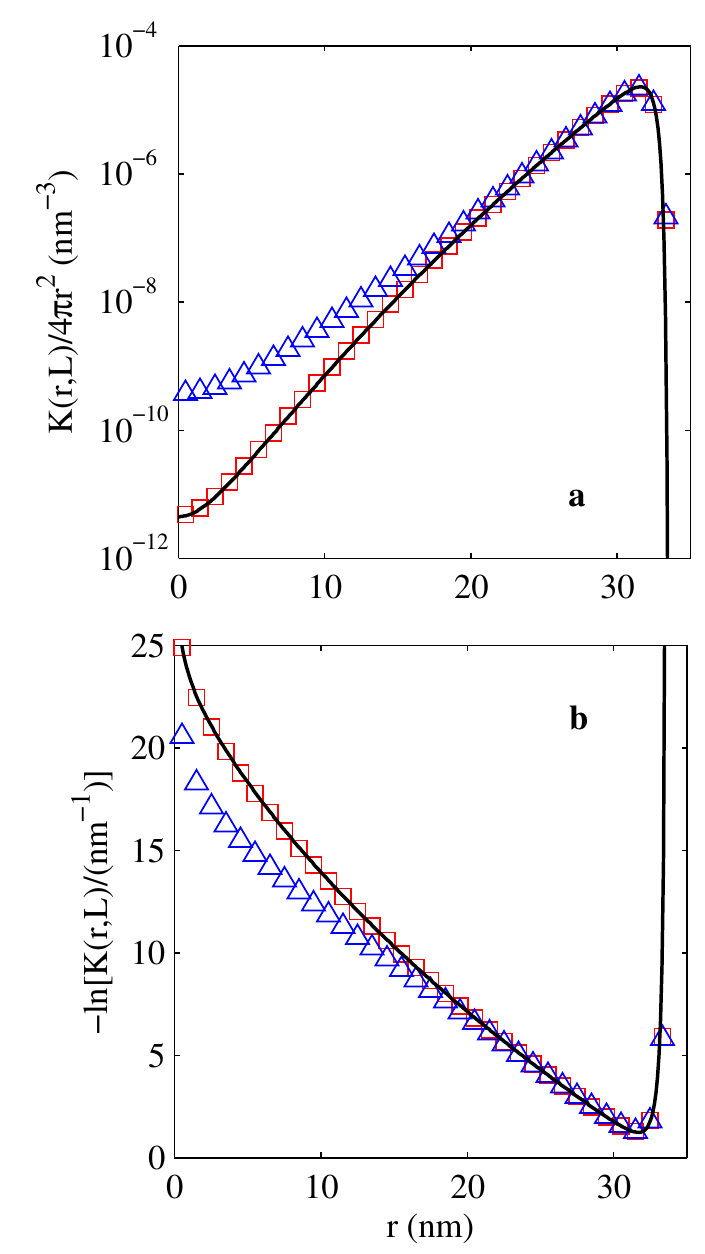}
\end{center}
\caption{(Color online) The radial distribution (a) and free energy (b) as a function of end-to-end distance, $r$, for a DNA of length $L=36.66\,\mathrm{nm}\,(=99\,\mathrm{bp})$. Data points represent MC simulation results, where triangles (blue) and squares (red) correspond to the models ``A" and ``W", respectively. Solid (black) curve corresponds to the theoretical prediction of the WLC model with $l_p=51\,\mathrm{nm}$~\cite{samuel}. Error bars (not shown) are about the size of the symboles. \label{fig:K}}
\end{figure}
\subsection{The end-to-end distribution functions}
Figures \ref{fig:K}(a) and \ref{fig:K}(b) compare the radial distribution function, $K(r,L)$ for a DNA of length $L=33.66\,\mathrm{nm}\,(=\,99\,\mathrm{bp})$. The triangles (blue) and squares (red) show MC simulation results for the models ``A" and ``W", respectively (Table \ref{table}). The solid (black) curve corresponds to the theoretical treatment of Samuel and Sinha~\cite{samuel} for the WLC model, which perfectly matches the simulation data. As can be seen in Figure \ref{fig:K}(a) there is no significant difference between the two models at large end-to-end distances, while at short end-to-end distances the radial distribution function in the AER model significantly deviates from that of the WLC model. 

\begin{figure}[thb]
\begin{center}
\centering
\subfigure{
\label{fig:6a}
\includegraphics[scale=0.45]{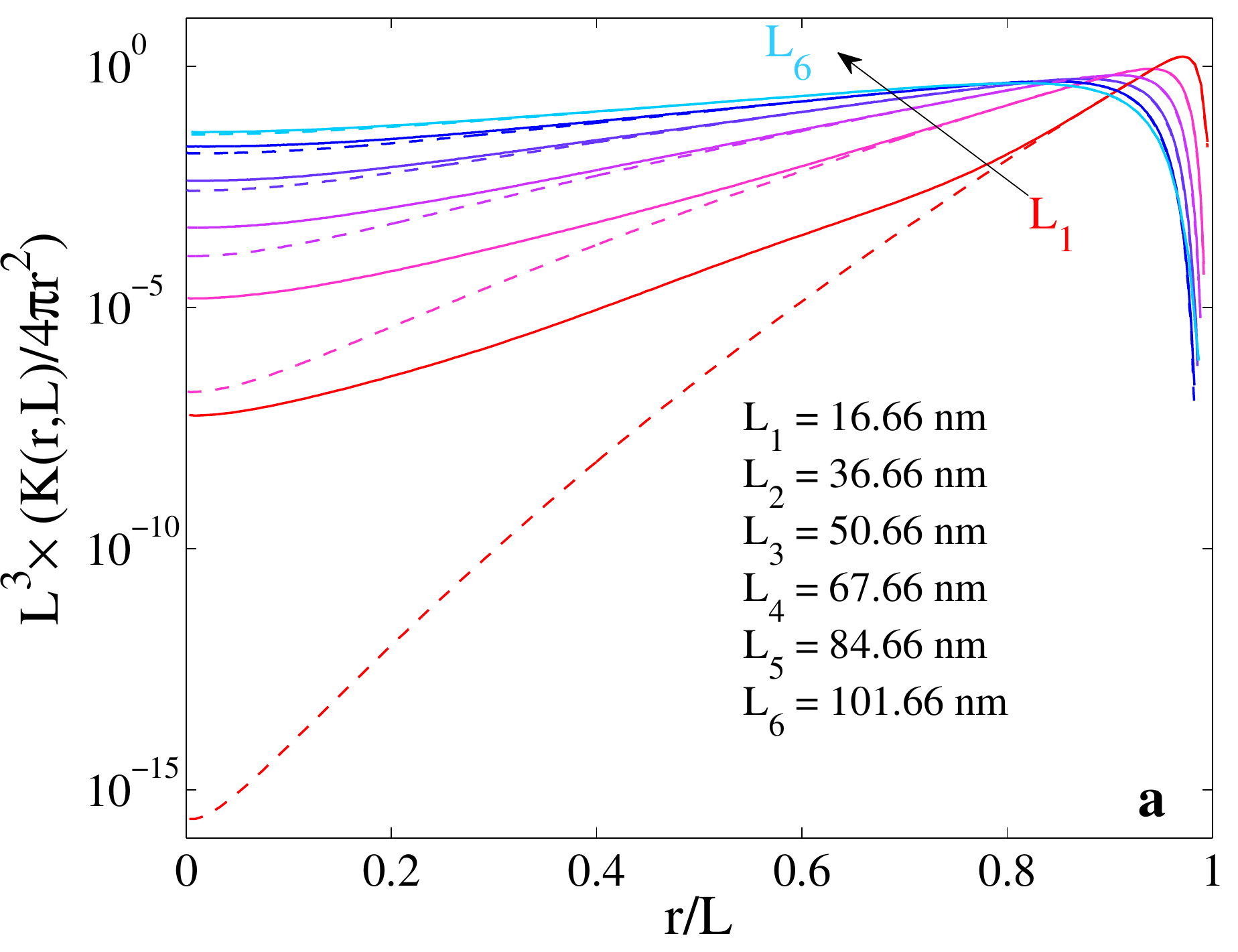}  
}\\
\subfigure{
\label{fig:6b}
\includegraphics[scale=0.47]{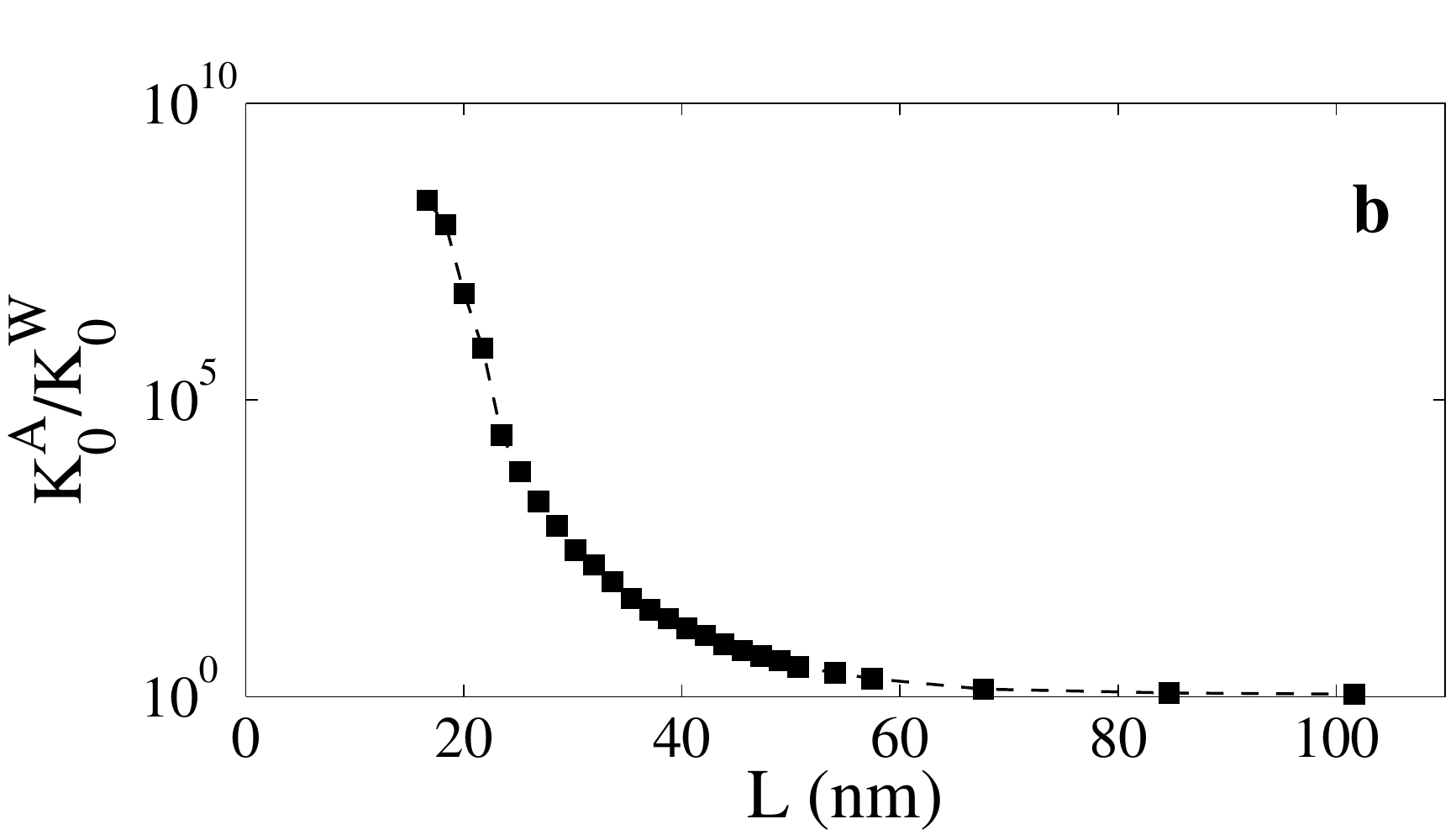}    
}
\end{center}
\caption{(Color online) a) $K(r,L)/4\pi r^2$ as a function of
$r$ for different DNA lengths. Solid and dashed curves corresponde to the models ``A" and ``W" in Table \ref{table}, respectively. b) $K^A_0/K^W_0$ as a function of DNA length. \label{fig:6}}
\end{figure}
Figure \ref{fig:K}(b) shows $-\ln(K(r,L))$, the free energy of the DNA as a function of its end-to-end distance. The position of the minimum in the free energy curve corresponds to the most probable end-to-end separation. A relaxed DNA molecule which is shorter than the persistence length tends to be almost straight. As can be seen in Figure \ref{fig:K}(b), for $L=99\,\mathrm{bp}$ the free energy minimum is very close to the total length of DNA. In this case we found that the average and the variance of the end-to-end distance in the WLC and AER models differ by less than $0.1$ percent. Therefore in the experiments which involve long free DNA molecules, such as DNA stretching experiment, the two models are indistinguishable. On the other hand, the asymmetric bending can significantly affect the outcome of the experiments performed on short, tightly bent DNA molecules, such as the DNA cyclization.

Figure
\ref{fig:6a} compares the distribution functions, $K(r,L)/4\pi r^2$, of the model ``A" (solid curves) with the model ``W" (dashed curve) for different lengths $L=16.66$, 36.66, 50.66, 67.66, 84.66, $101.66\,\mathrm{nm}$ (49, 99, 149, 199, 249 and 299 bp, respectively). One can see that the difference between the two models is disappeared as the DNA length increases. As expected for small end-to-end distances $K(r,L)/4\pi r^2$ converges to a constant $K_0(L)$ (see equation (\ref{greenfunc})) which is proportional to the unconstrained J-factor $J_0(L)$ (equation (\ref{J0factorG})). To calculate $K_0(L)$ we average $K(r,L)/4\pi r^2$ in vicinity of $r=0$ over the range of $0\leq r\leq r_0$, where $r_0$ is chosen such as $\frac{K(r_0,L)-4\pi r_0^2K_0(L)}{K(r_0,L)}\backsimeq0.01$. Figure \ref{fig:6b} shows the ratio $K_0^A/K_0^W$ as a function of DNA length, were the superscripts ``A" and ``W" refer to the AER and WLC models as parametrized in Table \ref{table}. As can be seen, while $K^A_0$ is several order of magnitude larger than $K^W_0$ for short DNA molecules, the ratio $K^A_0/K^W_0$ approaches unity as the DNA length increases.

\begin{figure}[thb]
\begin{center}
\subfigure{
\includegraphics[scale=0.20]{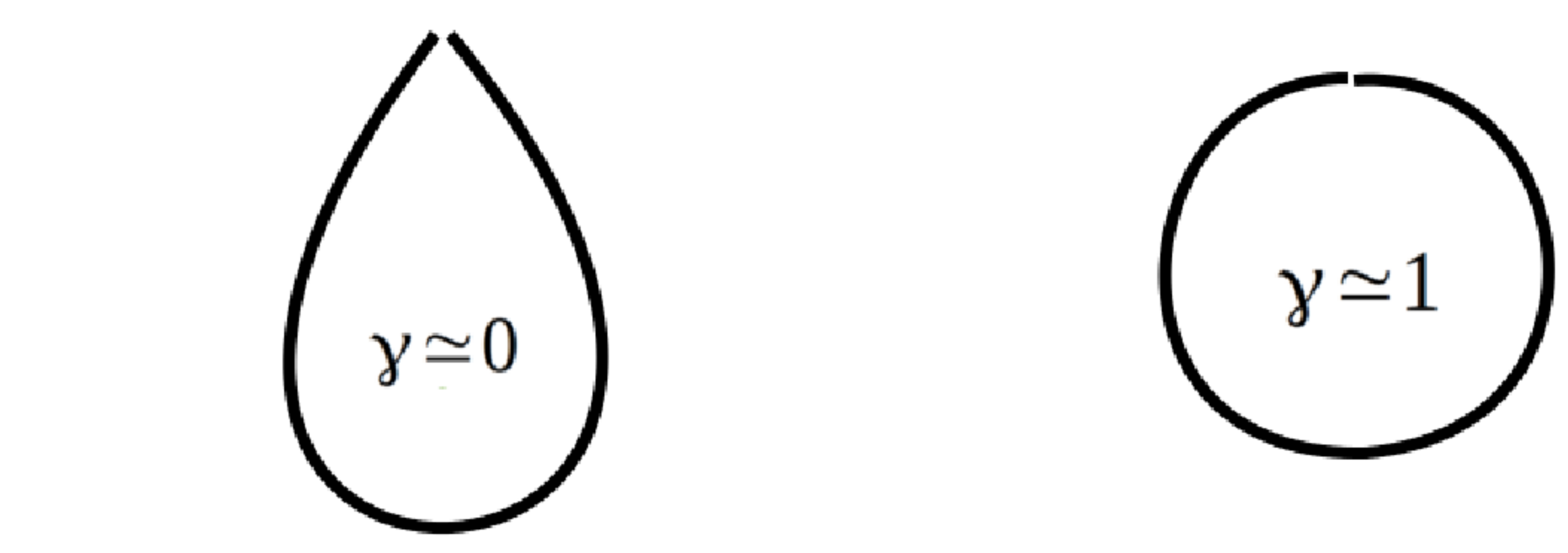}  
}\\
\subfigure{
\label{fig:15}
\includegraphics[scale=0.47]{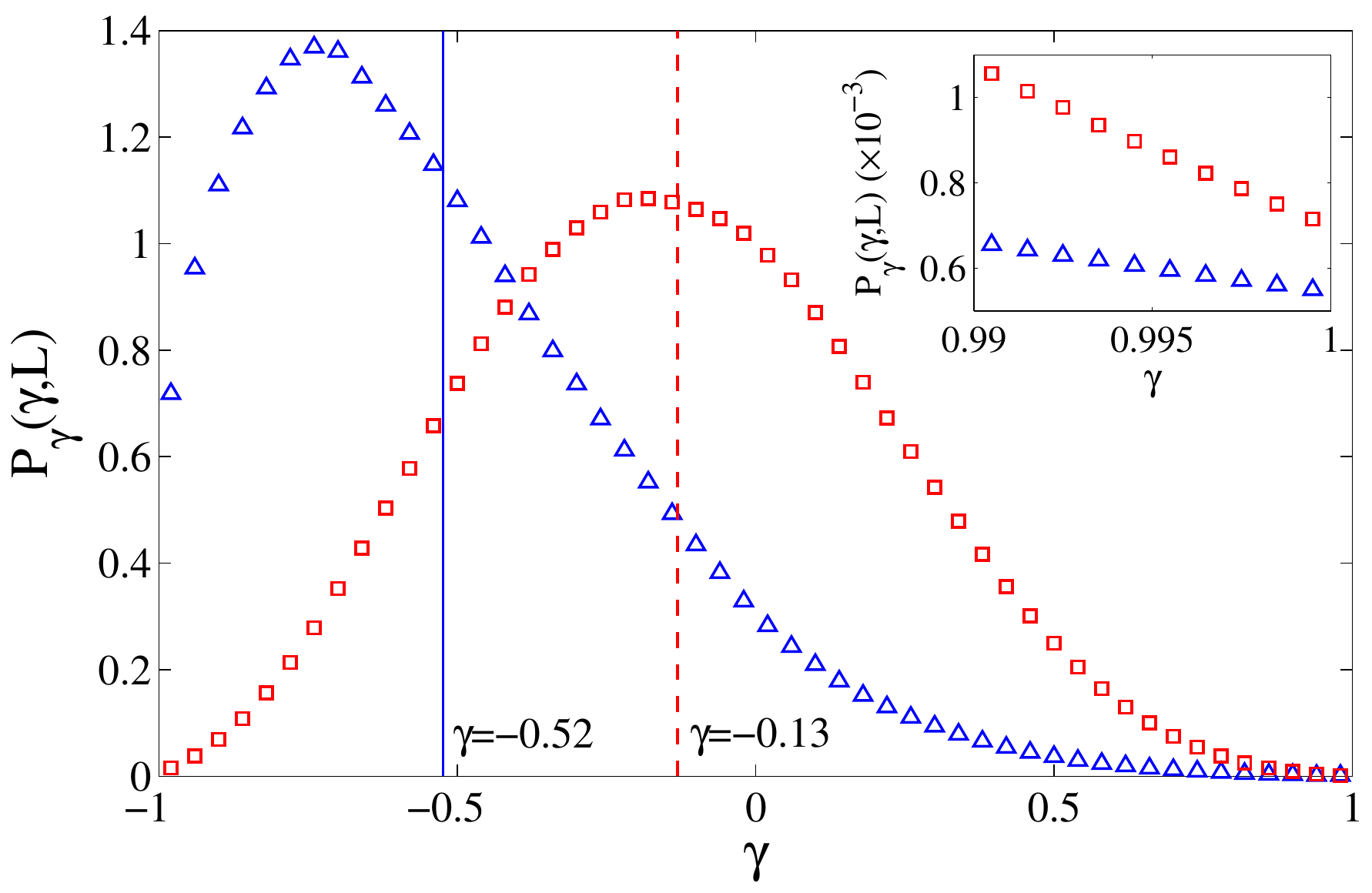}    
}
\end{center}
\caption{(Color online) Top: different schematic configurations of the chain for different angles between the two DNA tangents. Bottom: the MC simulation results for $P_{\gamma}(\gamma ,L)$ as a function of
$\gamma$ for a DNA loop of length $36.66\,\mathrm{nm}(=99\,\mathrm{bp})$. Triangles (blue) and squares (red) correspond to the models ``A" and ``W" in Table \ref{table} respectively. Solid (blue) and dashed (red) vertical lines (at $-0.52$ and $-0.13$, respectively) are the mean values of the distribution functions. Inset: show the tails of the distributions near $\gamma = 1$. Error bars (not shown) are about the size of the symboles. \label{fig:Pgamma}}
\end{figure}
\subsection{The Distribution Function of the End-to-end Tangent Vectors}
In Figure \ref{fig:Pgamma}, $P_{\gamma}(\gamma,L)$ is plotted against $\gamma$ for a DNA with $L=36.66\,\mathrm{nm}(=99\,\mathrm{bp})$ while its end-to-end distance is kept at $r=0$. In this Figure, triangles (blue) and squares (red) correspond to the models ``A" and ``W", respectively. As can be seen the two distributions are significantly different, but in the vicinity of $\gamma = 1$, they asymptotically approach each other. (inset of Figure \ref{fig:Pgamma}). We found that the peak of the distribution for the AER model generally occurs at a smaller $\gamma$ compared to the WLC model at short length scales (below the persistence length). For example, for $L=36.6\,\mathrm{nm}$, the most probable values of $\gamma$ for the models ``A" and ``W" are $-0.74$ and $-0.18$, respectively, and with $\langle \gamma\rangle^{A}=-0.52$ and $\langle \gamma\rangle^{W}=-0.13$ (as indicated in Figure \ref{fig:Pgamma} by solid (blue) and dashed (red) vertical lines, respectively). This indicates that when the two ends of a stiff chain meet each other, the angle between the terminal tangent vectors tends to be smaller in the AER model compared to the WLC model. This difference reflects the effect of the kink formation on the equilibrium structure of the DNA loop. The same structure also has been reported by other studies which are considered the kink in the model~\cite{Cocco2005,Kahn1998}.
\subsection{Loop Formation Probability}
\begin{figure}[thb]
\begin{center}
\includegraphics[scale=0.45]{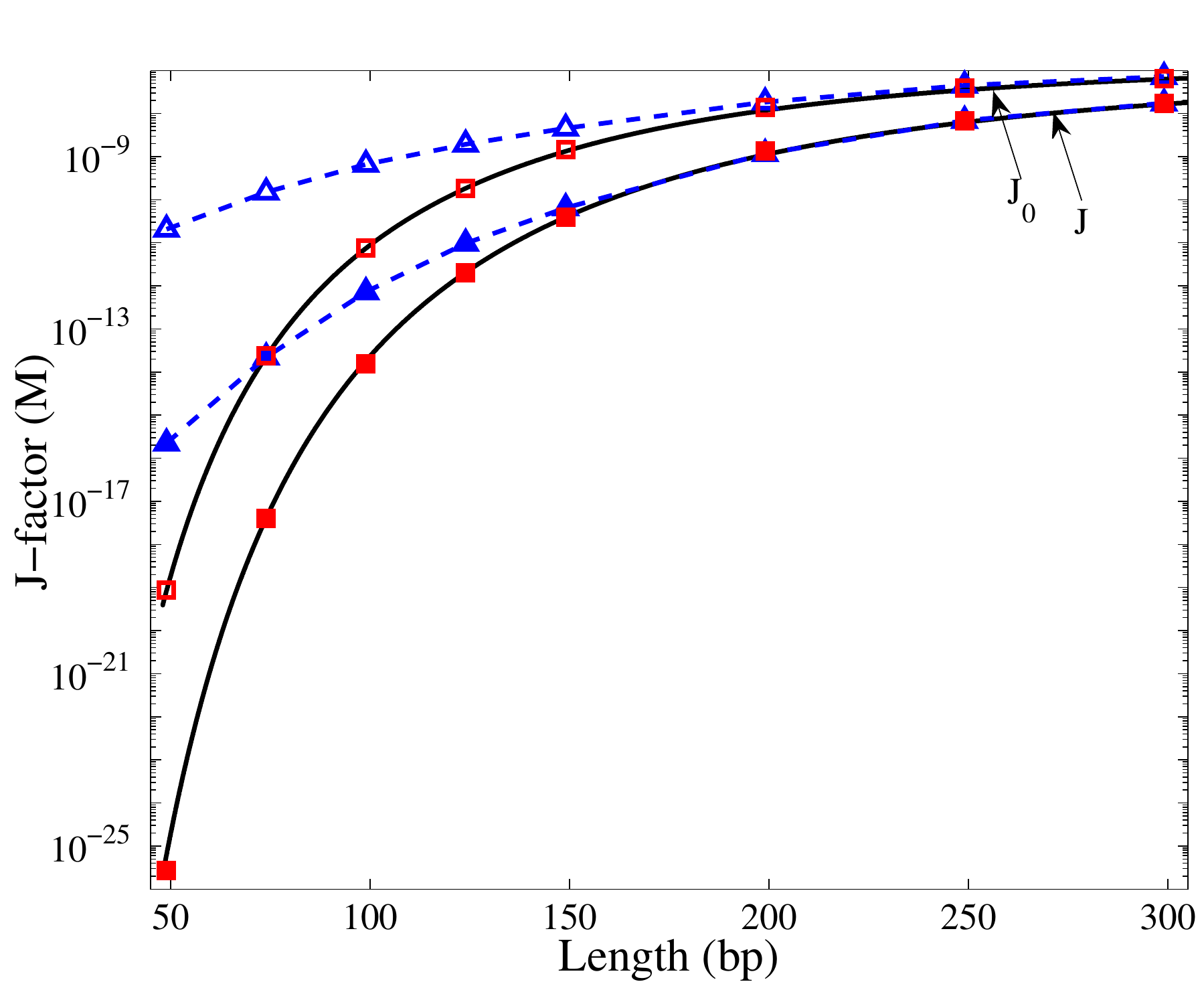}
\end{center}
\caption{(Color online) The J-factor as a function of DNA length. MC result for the model ``A" and ``W" are represented by triangles (blue) and squares (red), respectively, where filled and open symboles correspond to the cyclization with ($J$) and without ($J_0$) axial alignment. Shimada-Yamakawa's theoretical predictions for the J-factor in the model ``W" have been aslo shown by solid (black) curves~\cite{Shimada}. Dashed (blue) curves show the trend of the simulation data and do not correspond to a theoretical model. Error bars (not shown) are about the size of the symboles. \label{fig:loop}}
\end{figure}

Figure \ref{fig:loop} compares the the J-factor of the DNA in the AER (triangles, blue) and the WLC (squares, red) models, as obtained in the MC simulations, where filled and open symbols correspond to $J$ and $J_0$, respectively (see sections \ref{US} and \ref{sim}, and equations (\ref{JfactorG}) and (\ref{J0factorG})). The solid black curves, are the theoretical predictions for $J$ and $J_0$ in the WLC model~\cite{Shimada} which perfectly match the simulation data. In the case of the AER model, the dashed curves are shown as eye-guides. As can be seen, at short lengths (below $100\,\mathrm{bp}$), the J-factor in the AER model (with or with out axial alignment) is several orders of magnitude larger compared to the WLC model. As expected, the difference between the two models decreases as the DNA length increases, and for length larger than the DNA persistence length ($\sim150\,\mathrm{bp}$) the models are essentially indistinguishable. The same result can be obtained by other kinkable models~\cite{vologod2013,Wiggins02}. In this study we showed that the asymmetry in DNA structure may promote the kink formation, in particular, largely increases the J-factor at short length scales.
As it was discussed, the torsional constrain is not considered for the both looping probabilities $J$ and $J_0$, which leads to oscillations
of the J-factor as a function of DNA length with a period
equal to the DNA helical pitch~\cite{Shimada}.

\begin{figure}[thb]
\begin{center}
\includegraphics[scale=0.45]{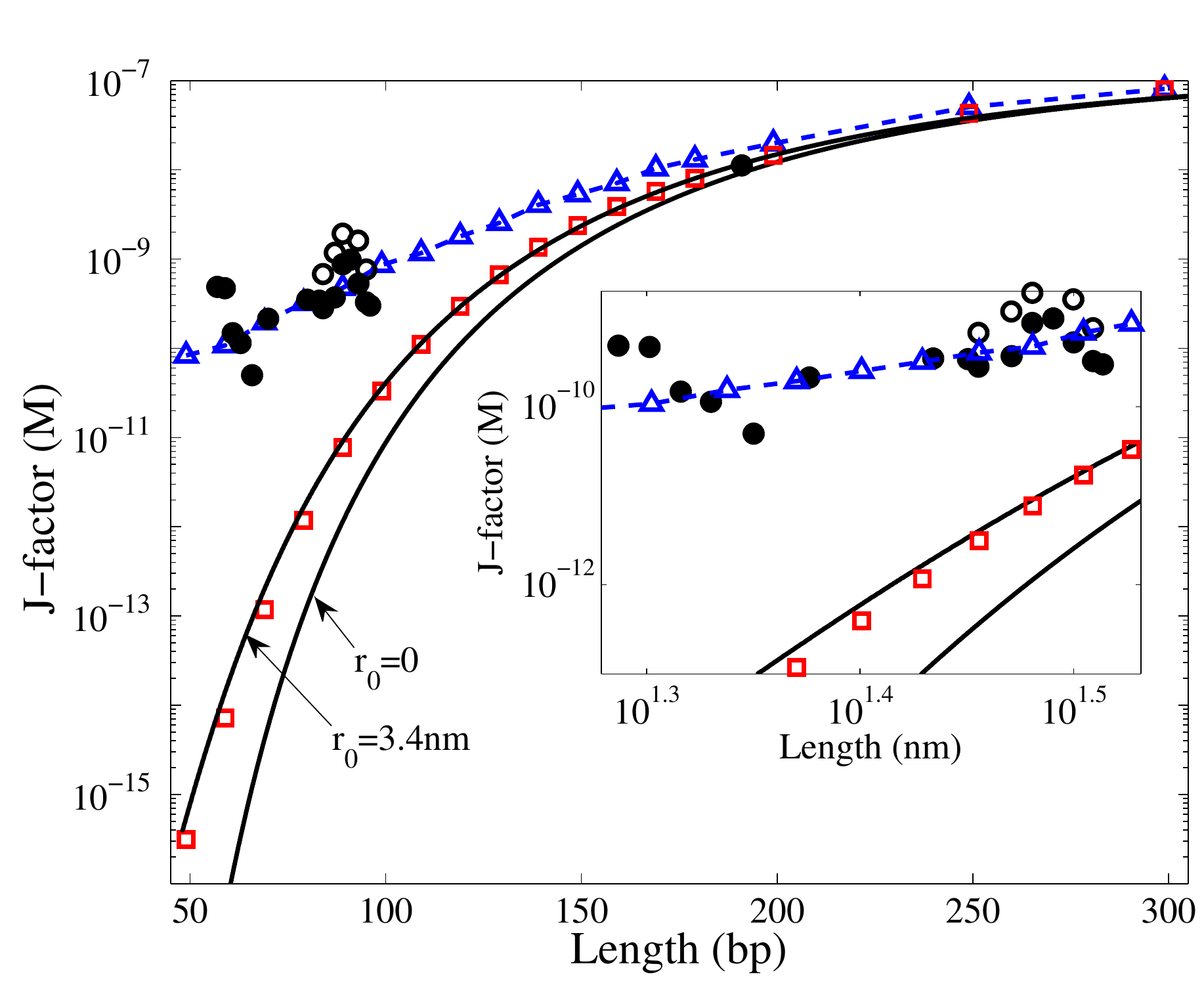}
\end{center}
\caption{(Color online) J-factor comparison. The triangles (blue) and squares (red) correspond to the simulation data of the model ``A" and ``W", respectively, for arbitrary angel between the duplex ends and the capture radius $r_0=3.4\,\mathrm{nm}$ ($=10\,\mathrm{bp}$). Circels show the experimental data of Vafabakhsh and Ha for surface tethered (filled) and vesicel encapsidated (open) experiments~\cite{Vafabakhsh}. For consistancy with our simulation data, we shifed the orginal data by $10\,\mathrm{bp}$ to the left for the sticky ends. The solid (black) curves are the theoretical predictions of the WLC model for $r_0=0$ and $3.4\,\mathrm{nm}$~\cite{Cocco2005}. Inset represents the length range of $50-100\,\mathrm{bp}$ in full logaritmic plot. Dashed (blue) curves show the trend of the simulation data and do not correspond to a theoretical model. Error bars (not shown) are about the size of the symboles. \label{fig:expjfac}}
\end{figure}

Recent experimental data of the J-factor of DNA molecules, which was performed by Vafabakhsh and Ha, have shown short DNA molecules are much more cyclizable than the prediction of the WLC model~\cite{Vafabakhsh}. In this experiment the DNA probe was a duplex with two complementary single-stranded overhangs on both ends (two sticky ends). Because the single-stranded overhangs are 10 nucleotide, they are considered as long sticky ends. It is expected that joining of such long sticky ends dose not require the axial alignment of the duplex ends~\cite{Vologodskii2013a}, and the effect of the torsional alignment could be considered as an oscillation factor in the J-factor. Also they can join each other when the end-to-end distance of the duplex is less than capture radius, $r_0$, which is $10\,\mathrm{bp}$ in this experiment. We thus evaluate the J-factor with free boundary condition and $3.4\,\mathrm{nm}$ capture radius for the both parameter models, i.e. the models ``A" and ``W". As Figure \ref{fig:expjfac} shows, while the experimental data significantly deviate from prediction of the WLC model for short DNA molecules, they show a considerable agreement with the AER model at all length scales. The oscillations in experimental data is believed to result from the torsional alignment between the DNA ends. In this issue, it has been suggested that underlying mechanism in the case of surface tethered may increase the rate of cyclization~\cite{Kim2013a,Kim2013b}, but this effect can not explain the anomalous behaviour of DNA. Other kinkable models show a sharp deviation from the WLC model at a critical length~\cite{Tung2014}, but the AER model shows a smooth deviation (see Figures \ref{fig:loop} and \ref{fig:expjfac}). The length dependence of the J-factor in the length range of $57$ to $96\,\mathrm{bp}$ is much weaker in the AER model compared to the WLC model. The inset of Figure \ref{fig:expjfac} shows a zoomed view in the length range of $50-100\,\mathrm{bp}$ in full logarithmic plot. To quantify the length dependence we fit power law functions to the simulations data points in this range and we found $J_0^A \sim L^{3.9}$ and $J_0^W \sim L^{16}$, where the superscripts indicate the model parameters ``A" and ``W".
\section{Conclusion}
In this paper, we proposed that the asymmetric structure of DNA can significantly affect the elasticity of DNA at short length scales. We have showed that, the extreme bendability of DNA at short lengths as well as the kink formation in double stranded DNA can originally form the asymmetric structure of DNA double helix. To account for the bending asymmetry we exploited the asymmetric elastic rod (AER) model, which has been introduced and parametrized in a previous study~\cite{Behrouz2}. By evaluating the effective bending energy and the distribution function of the end-to-end distance we show that although the AER model is equivalent to a WLC model at large length scales, for
tightly bent short DNA molecules the DNA is much more flexible in the AER model than in the WLC model. Using the umbrella sampling method, we evaluated the loop formation probability, i.e. the J-factor, as a function of the DNA length. We found that the unconstrained J-factor in the AER model with capture radius about $3.4\,\mathrm{nm}$ is in excellent agreement with the measured experimental data presented in~\cite{Vafabakhsh} at all length scales.
This implies that the axial alignment of the two ends is not required to join the two juxtaposed DNA ends in this experiment. Enforcing an axial alignment can induce an 1000-fold change in the J-factor (see Figure \ref{fig:Pgamma}). This may explain the large dispersity in the experimental data where DNA molecules with short sticky ends are cyclized~\cite{Widom01,Widom02,Vologodskii01}. The results presented in this paper, suggest that the asymmetric elastic rod model, as parametrized in~\cite{Behrouz2}, is a realistic model to explain the elastic behavior of DNA double helix at short length scales. 

\section{Acknowledgements}
We thank the Center of Excellence in Complex Systems and Condensed Matter (CSCM) for partial support.

\bibliography{biblio}

\end{document}